\documentclass[9pt,conference]{IEEEtran}
\IEEEoverridecommandlockouts

\usepackage[T1]{fontenc}
\usepackage{float} 
\usepackage{subfig}
\usepackage[export]{adjustbox} 
\usepackage{float} 
\usepackage{bbding}
\usepackage{booktabs}
\usepackage{multirow}
\usepackage{amssymb}
\usepackage{threeparttable}
\usepackage{stfloats}
\usepackage{xcolor}
\usepackage{pifont}
\usepackage{cite}
\usepackage{amsmath,amssymb,amsfonts}
\usepackage{algorithmic}
\usepackage{graphicx}
\usepackage{textcomp}
\usepackage{xcolor}
\usepackage[colorlinks=true, linkcolor=red, citecolor=black, urlcolor=magenta]{hyperref}
\def\BibTeX{{\rm B\kern-.05em{\sc i\kern-.025em b}\kern-.08em
    T\kern-.1667em\lower.7ex\hbox{E}\kern-.125emX}}
\begin{document}

\title{NeRF-3DTalker: Neural Radiance Field with 3D Prior Aided Audio Disentanglement for Talking Head Synthesis\\
\thanks{\IEEEauthorrefmark{1}Equal Contribution. \IEEEauthorrefmark{2}Corresponding Author:~\href{mailto:bichongke@tju.edu.cn}{bichongke@tju.edu.cn}.}
}

\author{\IEEEauthorblockN{Xiaoxing Liu\IEEEauthorrefmark{1}, Zhilei 
 Liu\IEEEauthorrefmark{1}, Chongke Bi\IEEEauthorrefmark{2}}
\IEEEauthorblockA{
\textit{College of Intelligence and Computing, Tianjin University, Tianjin, China}\\
}}

\maketitle

\begin{abstract}
Talking head synthesis is to synthesize a lip-synchronized talking head video using audio. Recently, the capability of NeRF to enhance the realism and texture details of synthesized talking heads has attracted the attention of researchers. However, most current NeRF methods based on audio are exclusively concerned with the rendering of frontal faces. These methods are unable to generate clear talking heads in novel views. Another prevalent challenge in current 3D talking head synthesis is the difficulty in aligning acoustic and visual spaces, which often results in suboptimal lip-syncing of the generated talking heads. To address these issues, we propose Neural Radiance Field with 3D Prior Aided Audio Disentanglement for Talking Head Synthesis (NeRF-3DTalker). Specifically, the proposed method employs 3D prior information to synthesize clear talking heads with free views. Additionally, we propose a 3D Prior Aided Audio Disentanglement module, which is designed to disentangle the audio into two distinct categories: features related to 3D awarded speech movements and features related to speaking style. Moreover, to reposition the generated frames that are distant from the speaker's motion space in the real space, we have devised a local-global Standardized Space. This method normalizes the irregular positions in the generated frames from both global and local semantic perspectives. Through comprehensive qualitative and quantitative experiments, it has been demonstrated that our NeRF-3DTalker outperforms state-of-the-art in synthesizing realistic talking head videos, exhibiting superior image quality and lip synchronization. Project page: \href{https://nerf-3dtalker.github.io/NeRF-3Dtalker/}{https://nerf-3dtalker.github.io/NeRF-3Dtalker/}.
\end{abstract}


\section{Introduction}
\vspace{-3pt}
Synthesizing talking head videos using audio is a popular research topic in the fields of multimedia and computer graphics. 
In the initial phases of research, numerous methods utilized image processing techniques or Generative Adversarial Networks (GAN) to generate image frames that match the input audio~\cite{ding2017exprgan, pumarola2018ganimation, ji2021audio, lu2021live, thies2020neural}. However, GANs are challenging to train \cite{chen2021talking}, and the lip shapes synthesized by these methods are often blurry and inaccurate. This can create bottlenecks in the field of talking head synthesis. 
Recently, researchers have investigated the use of diffusion models to synthesize talking head videos \cite{zhua2023audio, shen2022learning, stypulkowski2023diffused, multomm}. However, these models demand significant training resources and do not account for the spatial structure of the head, which restricts their ability to generate facial texture and 3D realism. 
In contrast, Neural Radiance Fields (NeRF) have gained attention for their ability to synthesize high-fidelity images with \textbf{limited datasets}. NeRF \cite{2020NeRF} is a novel technique for learning continuous volumetric representations of 3D scenes.
Although NeRF has been successfully applied to the talking head synthesis, most research focus on
rendering frontal faces
\cite{shen2022learning, bi2024nerf}. When substantial angular displacements of the head are encountered, the resultant images often exhibit a degradation in visual fidelity \cite{shen2022learning, bi2024nerf}. 

To tackle this challenge, we propose using 3D prior information from the 3D Morphable Model (3DMM) \cite{tran2018nonlinear} and semantics features to assist NeRF in generating talking heads with free views. 
The 3DMM model is a traditional method for extracting 3D prior information of faces, representing facial structure as identity, reflectance, and facial expression. 
Furthermore, current methods for synthesizing talking heads typically input the complete audio into the model \cite{yao2022dfa,shen2022learning,bi2024nerf}. However, the abundance of information in the audio, such as rhythm and timbre, can interfere with the model's learning process, reducing its ability to learn multimodal features.
To address this challenge, we propose a 3D Prior Aided Audio Disentanglement module to disentangle the input audio into two features: \(f_{exp-aud}\) and \(f_{exp-style}\). \(f_{exp-aud}\) represents the feature related to 3D awarded speech movements, while \(f_{exp-style}\) represents the feature related to speaking style. Speaking style refers to the movement of facial muscles influenced by audio. The 3D Prior Aided Audio Disentanglement module \textbf{aligns the acoustic space with the 3D visual space}, which is crucial in the field of 3D talking head synthesis. Subsequently, we input \(f_{exp-aud}\) and \(f_{exp-style}\) along with the 3D prior information obtained from the 3DMM model into the presented conditional NeRF to synthesize talking heads with free views.
Furthermore, 
in order to bring back the generated frames that are distant from the speaker's motion space to the real space, we draw inspiration from the concept of codebooks \cite{van2017neural} and design Standardized Space for speakers. Specifically, we model the speaker's motion space as a global Standardized Space. 
After that, we input the generated frames into this Standardized Space to normalize the irregular positions in the generated frames. 
Additionally, due to the strong correlation between the speaker's motion and speech movements, we propose the use of a semantic standardized space to express local semantics. This will effectively constrain the generated heads from both global and local perspectives. 
To summarize, this paper presents the following contributions:
\begin{itemize}
\item We provide audio semantics and visual 3D prior features for conditional NeRF to synthesize reliable multi-view talking heads.
\item 
To align the acoustic space with the 3D visual space, 
we propose a 3D Prior Aided Audio Disentanglement module that disentangles the audio into features related to 3D awarded speech movements and features related to speaking style.
\item We design a Standardized Space that improves the realism and regularity of the generated images from both global and semantic perspectives.
\end{itemize}

\begin{figure*}
  \centering
  \includegraphics[width=0.9\textwidth]{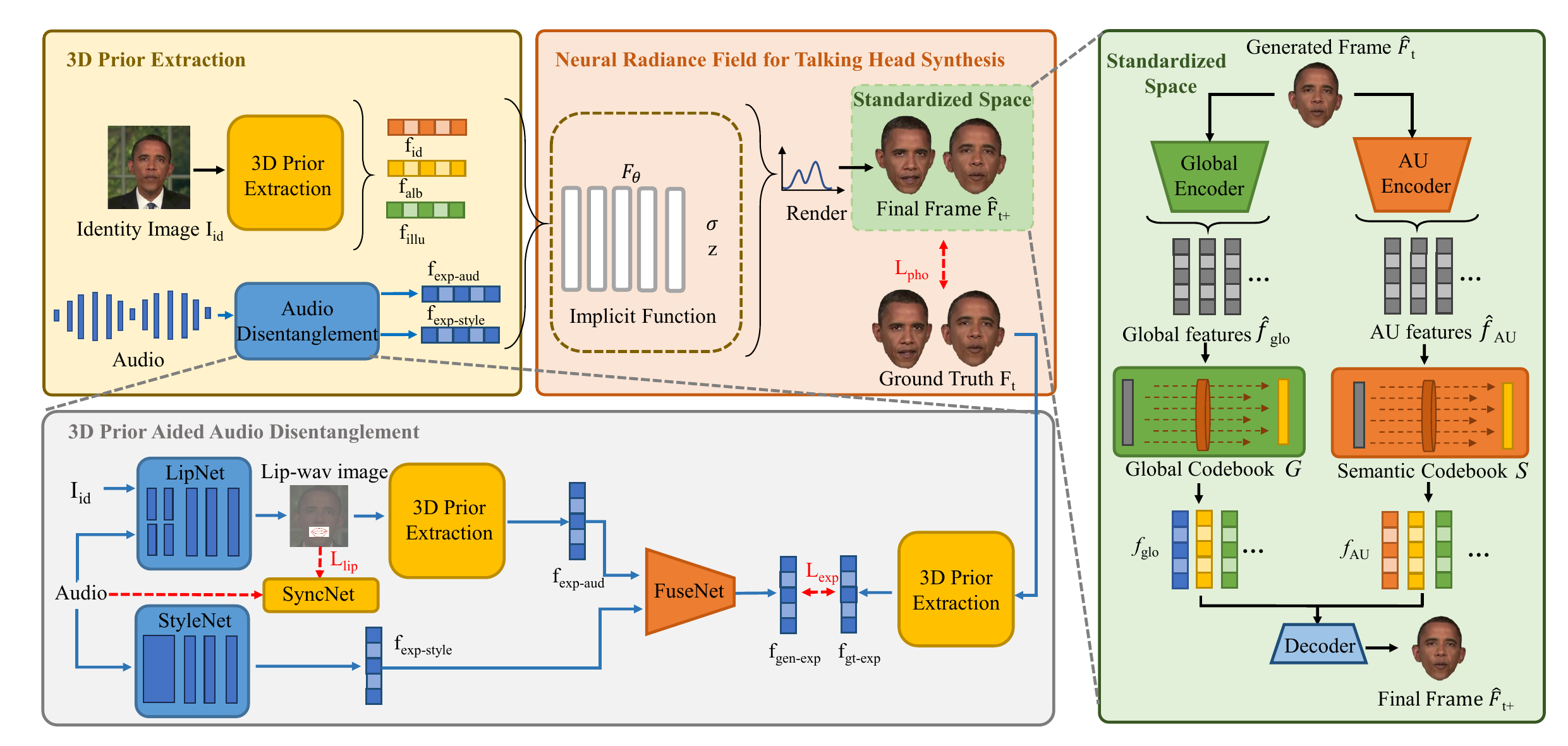}
  \caption{Overview of NeRF-3DTalker. 
  }
    \vspace{-16pt}
  \label{fig:teaser2}
\end{figure*}

\vspace{-8pt}
\section{Proposed method}
\vspace{-4pt}
The overall architecture of our proposed NeRF-3DTalker is shown in Fig.\ref{fig:teaser2}, which consists of four main components: the 3D Prior Extraction module (enclosed by the light yellow rectangle), the 3D Prior Aided Audio Disentanglement module (enclosed by the gray rectangle), the Neural Radiance Field Rendering module (enclosed by the orange rectangle), and the Standardized Space (enclosed by the green dashed rectangle). In the following sections, we provide detailed explanations of each component.

\subsection{3D Prior Extraction}
To generate a talking head with free views, we incorporate 3D facial priors into our model with the assistance of 3DMM. 
Inspired by \cite{egger20203d, tran2018nonlinear, hong2022headnerf}, we parameterize a head into shape and appearance. For appearance, similar to 3DMM, we use facial albedo \(f_{alb}\), and scene illumination \(f_{illu}\) to control.
In order to better reconstruct the 3D talking head, we use identity \(f_{id}\), expression related to 3D awarded speech movements \(f_{exp-aud}\), expression related to speaking style \(f_{exp-style}\) for shape representation, as shown in Eq.~\ref{Eq1}:

\begin{equation}
\label{Eq1}
    I:(f_{id},f_{exp-aud},f_{exp-style},f_{alb},f_{illu})
\end{equation}
where \(f_{id}~\in~\mathbb{R}^{100}\), \(f_{exp-aud}~\in~\mathbb{R}^{79}\), \(f_{exp-style}~\in~\mathbb{R}^{79}\), \(f_{alb}~\in~\mathbb{R}^{100}\), \(f_{illu}~\in~\mathbb{R}^{27}\) and \(I\) represents the speaker's head. 
\(f_{exp-aud}\) and \(f_{exp-style}\) are obtained through 3D Prior Aided Audio Disentanglement Module.

\subsection{3D Prior Aided Audio Disentanglement}
Since speech audio contains a lot of mixed information, including rhythm and timbre, feeding the complete audio into the NeRF would increase its learning burden and fail to learn precise speech motion features. To address this issue and align
acoustic space with 3D visual space, we propose a 3D Prior Aided Audio Disentanglement module that disentangles the speech audio into features related to 3D awarded speech movements \(f_{exp-aud}\) and features related to speaking style \(f_{exp-style}\).

As shown in the gray area of Fig.\ref{fig:teaser2}, the first step is to input the identity image \(I_{id}\) and audio into a pre-trained LipNet network \cite{zhang2022sadtalker} to generate Lip-wav images focused on the motion of the lips. 
To accomplish this, inspired by \cite{prajwal2020lip}, we also use a pre-trained SyncNet to provide lip synchronization loss for supervision. The SyncNet we use judges whether the generated images and speech are synchronized. It consists of a face encoder and an audio encoder. The lip synchronization loss provided by SyncNet can be represented as follows:
\begin{equation}
    L_{\mathrm{lip}}=\frac{1}{B}\sum_{i=1}^{B}-\log(\frac{f_{lip_i}\cdot f_{a_i}}{max(\|f_{lip_i}\|_2\cdot\|f_{a_i}\|_2,\epsilon)})
\end{equation}
where $B$ denotes the number of frames processed per iteration. \(f_{lip_i}\) denotes the feature obtained by inputting the generated Lip-wav image into the face encoder, and \(f_{a_i}\) represents the feature obtained by inputting the audio sequence into the audio encoder.

Then, we use 3D Prior Extraction module to extract the expression features \(f_{exp-aud}\) from the Lip-wav images. These expression features are only related to the speech content.
And we use a StyleNet network to extract features related to the individual's speaking style. Next, we input these two features into the fusion module (FuseNet) to generate fused expression features. We calculate the loss between the fused features \(f_{gen-exp}\) and the expression features \(f_{gt-exp}\) of the real image to optimize the entire 3D Prior Aided Audio Disentanglement module. The loss is as follows:
\begin{equation}
    L_{\mathrm{exp}}=||f_{gen-exp}-f_{gt-exp}||^2
\end{equation}

\begin{figure}[b]
  \centering
  \includegraphics[width=0.9\linewidth]{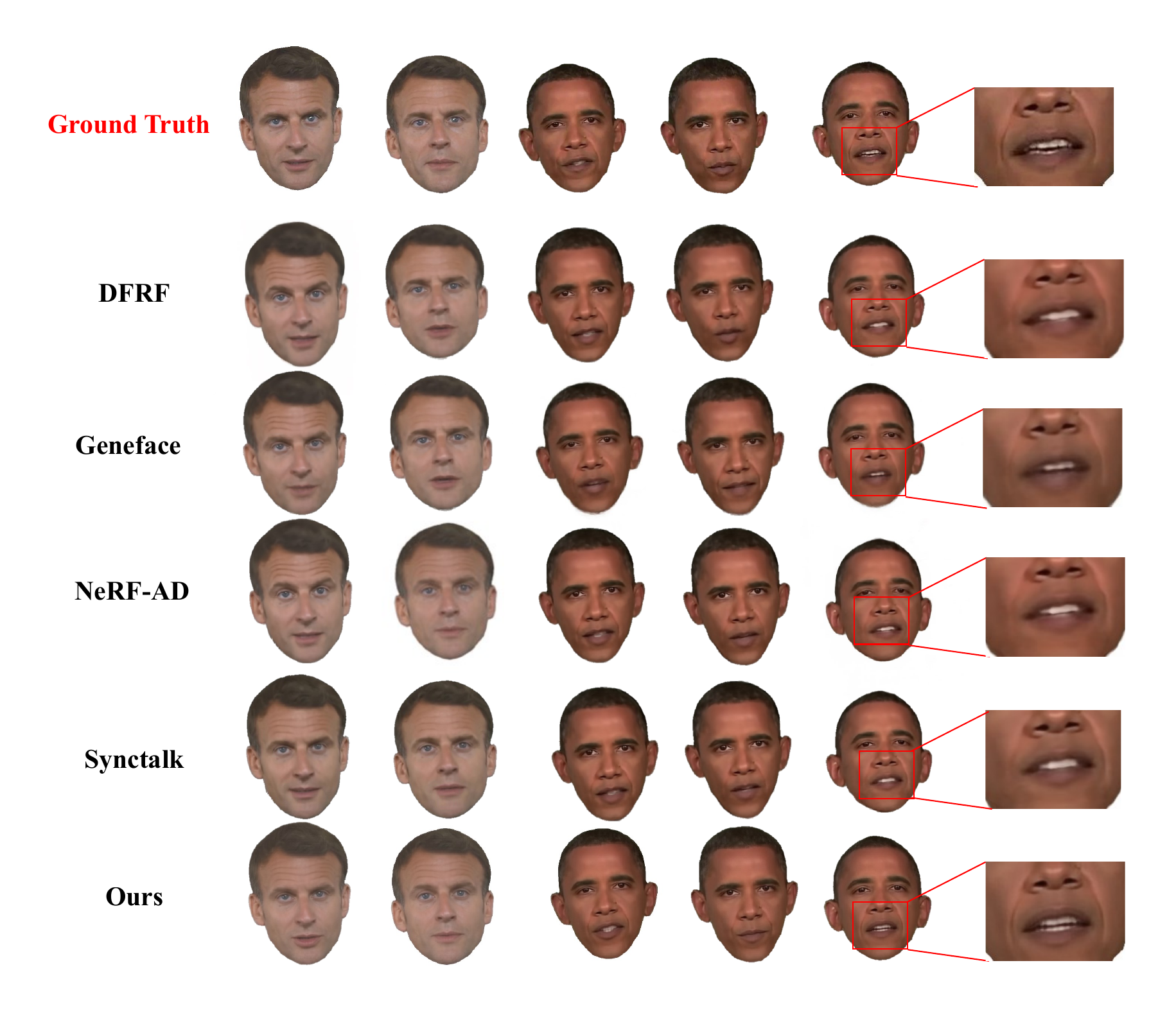}
  \caption{Visual results of the comparative experiments.}
  \label{fig:teaser3}
\end{figure}

\begin{figure}[b]
  \centering
  \includegraphics[width=\linewidth]{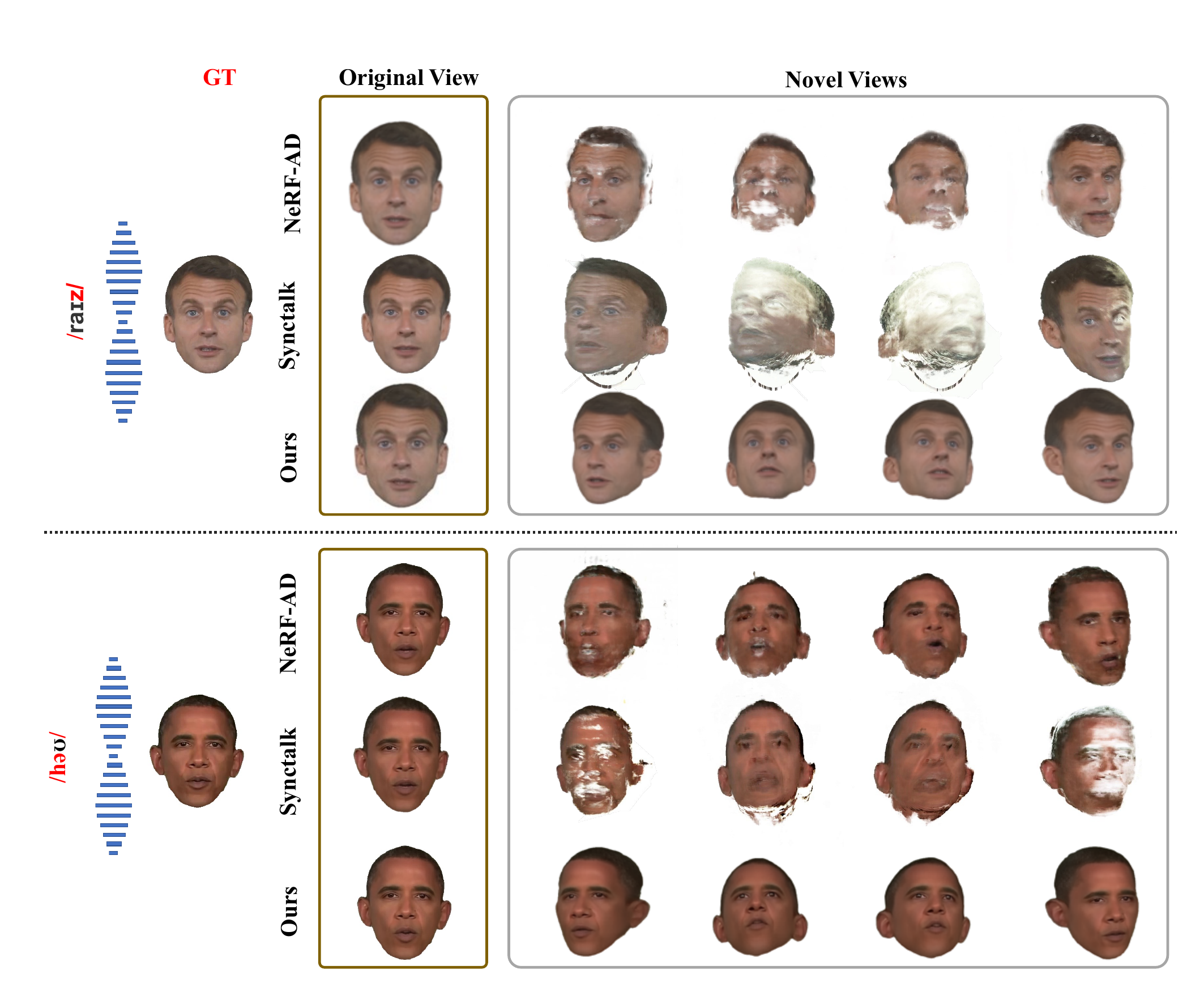}
  \caption{Novel view results of the comparative experiments.}
  \label{fig:teaser4}
\end{figure}

\subsection{NeRF for Talking Head Synthesis}
\label{Sec_NeRF}
Inspired by the work of \cite{shen2022learning}, we present a conditional NeRF to synthesize the talking head images with \(f_{id}\), \(f_{exp-aud}\), \(f_{exp-style}\), \(f_{alb}\), \(f_{illu}\) as conditions. 
We then use these conditions along with the 3D position \(l=(x,y,z)\) and camera parameter \(P\) as the input to the implicit function \(F_\theta\).
Finally, the NeRF, consisting of multi-layer perceptrons, can estimate a high-dimensional feature vector \(z\) and density \(\sigma\) of each 3D point on each ray. The entire implicit function can be formulated as follows:
\begin{equation}
    F_\theta:(l,P,f_{id},f_{exp-aud},f_{exp-style},f_{alb},f_{illu})\longrightarrow(z,\sigma)
\end{equation}
To generate the output image \(\hat{F}_t\), volume rendering is performed by integrating the high-dimensional feature vectors and densities of all 3D points along each ray \(r\). This process is represented as:
\[\hat{F}_t(r;\theta,f_{id},f_{exp-aud},f_{exp-style},f_{alb},f_{illu})=\]
\begin{equation}
    H_u(\int_{t_n}^{t_f}\sigma(t)\cdot z(t)\cdot T(t)dt)
\end{equation}
where \(t_n\) and \(t_f\) denote the near and far bound of a camera ray respectively. \(T(t)=exp(-\int_{t_n}^{t}\sigma(r(\tau))d\tau)\) denotes the integral transmittance along a camera ray from near \(t_n\) bound to $t$. 
It is important to note that there is significant pressure on the rendering of Neural Radiance Field due to the requirement of generating multi-view talking heads. To decrease the rendering time of NeRF, we employ the upsampling operation \(H_u\), following the inspiration of~\cite{niemeyer2021giraffe,karras2020analyzing}. 

The loss function \(L_{pho}\) is utilized in this process to minimize the photo-metric reconstruction error between the generated image \(\hat{F}_t\) and the ground truth \(F_t\).
\begin{equation}
    L_{pho}=||\hat{F}_t(r;\theta,f_{id},f_{exp-aud},f_{exp-style},f_{alb},f_{illu})-F_t||^2
\end{equation}
where \(\theta\) denotes the model parameters.

\subsection{Local-Global Standardized Space}
\label{ss}

Due to the complexity of the 3D talking head synthesis task, current methods inevitably produce some generated frames that deviate from the speaker’s motion space, which limits the visual effect of the original view.
In order to bring back the generated frames that are distant from the speaker's motion space to the real space, we draw inspiration from the concept of codebooks \cite{van2017neural} and design a Standardized Space for speakers.

The Standardized Space we designed comprises a \textbf{global space} and a \textbf{local AU semantic space}, as illustrated in Fig.\ref{fig:teaser2}. The local AU semantic space enhances the semantic information of the generated frames from the perspective of the local AU, while the global space improves the global information of the generated frames from the perspective of the overall face. The semantic information refers to the features in the face that are highly correlated with speech movements. 
In this paper, we use speech-related Action Units (AUs) information to represent semantic elements. 

The local AU semantic space comprises an AU encoder and a semantic codebook $S$. The main objective during training is to learn how to establish the semantic codebook $S$ that stores the standardized semantic elements. This enables the replacement of semantic elements in generated frames with standardized ones using Eq.~\eqref{eq1} during testing.

We first input the real frames into a pre-trained AU encoder to generate AU features.
The AU encoder comprises four convolutional layers and four fully connected layers. 
To train the encoder, we use binary cross-entropy loss to calculate the AU distance between the input frames and the ground truth. The specific loss function is shown below:
\begin{equation}\label{bce}
    L_{bce}=-\frac{1}{n_{AU}}\sum_{i=1}^{n_{AU}}w_i[x_{i}log\hat{x}_{i}+(1-x_{i})log(1-\hat{x}_{i})]
\end{equation}
where \(x_i\) and \(\hat{x}_{i}\) denote the $i$-th ground truth AU label and the $i$-th input frame AU label respectively. The above AU labels are extracted using OpenFace \cite{baltruvsaitis2015cross,baltrusaitis2018openface}. To reduce the impact of the correlation between AUs on model training, we adopt the method proposed by \cite{chen2021talking}, which adds weights \(w_i=\frac{\frac{1}{r_i}n_{AU}}{\sum_{i=1}^{n_{AU}}\frac{1}{r_i}}\) to Eq.~\eqref{bce}, where \(r_i\) is the probability of the $i$-th AU occurring in the dataset. Meanwhile, since some AUs appear rarely in the training set and can affect the model's prediction results, we introduce a weighted multi-label Dice coefficient loss \cite{chen2021talking, milletari2016v}.
\begin{equation}
    L_{dice}=\frac{1}{n_{AU}}\sum_{i=0}^{n_{AU}}w_i[1-\frac{2x_i\hat{x}_{i}+\varepsilon}{x_i^2+\hat{x}_{i}^2+\varepsilon}]
\end{equation}
Our total AU loss is \(L_{AU}=L_{bce}+L_{dice}\).


After obtaining the AU features \(\hat{f}_{AU}\), they are inputted into the codebook $S$. The query function $Q$ is then used to obtain the feature in the codebook $S$ that is closest to the input. The output features are the standardized AU features \(f_{AU}\). This process is defined as:
\begin{equation}
    f_{AU}=Q(\hat{f}_{AU}):=\underset{f_{{AU}_i}\in\mathcal{S}}{\arg\min}\|\hat{f}_{{AU}_i}-{f_{AU}}_i\|_2
    \label{eq1}
\end{equation}

Similar to the local AU semantic space, the global space comprises a global encoder and a global codebook $G$. The face is first inputted into the global encoder \cite{xing2023codetalker} to generate global facial features \(\hat{f}_{glo}\). \(\hat{f}_{glo}\) is then inputted into the global codebook $G$ to query and obtain standardized global features \(f_{glo}\). This process is defined as:
\begin{equation}
    f_{glo}=Q(\hat{f}_{glo}):=\underset{f_{{glo}_i}\in\mathcal{G}}{\arg\min}\|\hat{f}_{{glo}_i}-{f_{glo}}_i\|_2
    \label{eq2}
\end{equation}

Finally, the features obtained from both the local AU semantic space and the global space are inputted into the Decoder to generate the final image \(\hat{F}_{t+}\).

Constraints are imposed on the training process in standardized space from two perspectives: the angle of the generated frames and the angle of the codebook features. The loss function is shown as follows:
\[L_{S}=\|\hat{F}_{t+}-F_{t}\|_{1}\\+\|\mathrm{sg}(\hat{f}_{AU})-f_{AU}\|_{2}^{2}+\]
\begin{equation}
    \beta_1\|\hat{f}_{AU}-\mathrm{sg}(f_{AU})\|_{2}^{2}+\|\mathrm{sg}(\hat{f}_{glo})-f_{glo}\|_{2}^{2}+\beta_2\|\hat{f}_{glo}-\mathrm{sg}(f_{glo})\|_{2}^{2}
    \label{eq4}
\end{equation}
where $sg$ denotes a stop-gradient operation. This operation is defined as an identity function during forward propagation, but its gradients are not computed during backpropagation. This effectively constrains its operand to remain a non-updated constant. And \(\beta_1\) and \(\beta_2\) denote the trade-off parameters.


\begin{table*}[]
\resizebox{\textwidth}{!}{
\begin{tabular}{c|ccccc|ccccc|c}
\toprule
\multirow{2}{*}{Method} & \multicolumn{5}{c|}{TestA} & \multicolumn{5}{c|}{TestB} & \multirow{2}{*}{\textcolor{blue}{Ne}} \\ \cline{2-11}
                & SSIM↑  & LPIPS↓ & AU Acc↑ & LMD-79↓  & SyncNet↑   & SSIM↑   & LPIPS↓  & AU Acc↑        & LMD-79↓  & SyncNet↑   &      \\ \midrule \midrule
    Ground-truth  & -      & -      & -       & -        & 2.729      & -                   &-        &-               &-         &2.772       & -     \\
    Wav2Lip 2020\cite{prajwal2020lip}  &  0.793 &0.052    & 72.2\%    & 2.921      & \textbf{2.237}   &0.821 &0.059   &76.7\%  &3.067  &\textbf{1.971} & \ding{53} \\
    DFRF 2022\cite{shen2022learning}  & 0.818  & 0.045 & 71.4\% & 3.003 & 1.765     & 0.851  & 0.041  & 75.7\% & 2.859  & 1.694  & \ding{51}                \\
    GeneFace 2023\cite{ye2023geneface}   & 0.811 & 0.042  & 73.1\%   & 2.815  & 1.793   & 0.857  & 0.036 & 75.4\%  & 2.864 & 1.712  & \ding{51} \\ 
    NeRF-AD 2024\cite{bi2024nerf}   & 0.831 & 0.040  & 73.3\%   & 2.633  & 1.801   & 0.882  & 0.032 & 76.9\%  & 2.772 & 1.756  & \ding{51} \\ 
    Synctalk 2024\cite{synctalk}   & 0.828 & 0.043  & 73.6\%   & \textbf{2.511}  & 1.793   & \textbf{0.885}  & 0.029 & 76.9\%  & 2.852 & 1.727  & \ding{51} \\ \midrule
    \textbf{Ours}  & \textbf{0.835} & \textbf{0.037} & \textbf{75.3\%} & 2.529 & 1.956  & 0.880 & \textbf{0.025} & \textbf{78.4\%} & \textbf{2.650} & 1.868& \ding{51}  \\ \bottomrule
\end{tabular}}

\caption{Quantitative results compared with other methods. Best results are in \textbf{bold}. \textcolor{blue}{"Ne"} indicates whether this method is based on NeRF.}
\vspace{-10pt}
\label{tab:t1}
\end{table*}

\section{Experiments}
\subsection{Experimental Settings}
\textbf{Dataset.}
To compare with state-of-the-art methods, follow the approach of \cite{shen2022learning, ye2023geneface}, we collect four videos of different individuals giving speeches, interviews, or reading news from their publicly released video sets. 
Additionally, we preserve the facial head segment within the dataset to facilitate the generation of 3D talking heads. Following the approach of \cite{yao2022dfa}, these videos are randomly divided into two groups for experimentation.

\textbf{Evaluation Metrics.}
The quality of the generated images is evaluated using SSIM \cite{wang2004image}, and LPIPS ~\cite{zhang2018unreasonable}. Lip synchronization is evaluated using AU Acc \cite{chen2021talking}, landmark distance (LMD)~\cite{chen2018lip}, and SyncNet confidence value \cite{prajwal2020lip}. In order to evaluate the lip shape more accurately, we have improved the LMD metric by increasing the number of calculated lip landmarks from 20 to 79, consistent with the approach of \cite{bi2024nerf}. 



\subsection{Quantitative Results}

The experimental results are shown in Table~\ref{tab:t1}. The table shows that our method outperforms the others in most of the metrics. 
Our method achieves a decrease of $0.016$ in LPIPS↓ and a significant improvement of 2.7\% in AU Acc↑ compared to DFRF \cite{shen2022learning} on dataset B. In terms of lip synchronization metrics, our method achieves an increase of 3\% in AU Acc↑ and a decrease of 0.021 in LMD-79↓ compared to Geneface \cite{ye2023geneface} on dataset B. 
Our method also increases AU Acc↑ by 1.7\% on Test A and by 1.5\% on Test B compared with Synctalk \cite{synctalk}. 
Overall, compared with NeRF-based methods, our method achieves the highest level of lip synchronization, demonstrating the importance of aligning acoustic space with 3D visual space in 3D talking head synthesis.
We also compare our method with non-NeRF-based method. Our method demonstrates significant advancements in image quality compared to Wav2Lip \cite{prajwal2020lip}. Regarding lip synchronization, since Wav2Lip was trained on a large dataset specifically for this metric, it slightly outperforms ours on the SyncNet metric. 

\begin{table}[b]
\resizebox{\linewidth}{!}{
\begin{tabular}{cccccc}
\toprule
\multicolumn{1}{c|}{Method}     & SSIM↑          & LPIPS↓         & AU Acc↑        & LMD-79↓           & SyncNet↑       \\ \midrule \midrule
\multicolumn{1}{c|}{Ground-truth}      & -              & -              & -              & -              & 2.729          \\ 
\multicolumn{1}{c|}{use ent. aud.}       & 0.796          & 0.046          & 69.1\%          & 3.001          & 1.412          \\
\multicolumn{1}{c|}{use spe. aud.}                & 0.808          & 0.044 & \textbf{72.7\%}          & 2.794          & 1.699          \\
\multicolumn{1}{c|}{use dis. aud.}         & \textbf{0.819}          & \textbf{0.041}          & 72.4\%          & \textbf{2.637}          & \textbf{1.721}          \\  \bottomrule
\end{tabular}}
\caption{The ablation study of 3D Prior Aided Audio Disentanglement module.}
\label{tab:t2}
\end{table}


\subsection{Qualitative Results}
The qualitative comparison results with other methods are shown in Fig.\ref{fig:teaser3}. It is evident that our method produces mouth images that are closest to the Ground Truth, as indicated by the red region in the graph. 
In addition, by adjusting the camera poses of NeRF, we also compare our NeRF with the other NeRF models for multi-view synthesis.
The comparison results are depicted in Fig.~\ref{fig:teaser4}.
The results demonstrate that our method generates the best images under identical camera pose conditions.
This is due to datasets containing only frontal views of talking heads, which limits the ability of NeRF to accurately reconstruct a comprehensive three-dimensional space for talking heads.
In contrast, our approach integrates 3D prior knowledge and a 3D Prior Aided Audio Disentanglement module, enabling NeRF to reconstruct crisp 3D talking head features.

\subsection{Ablation Studies}

To measure the impact of the primary components of our proposed model, we conduct ablation studies, which are divided into two parts.

Table \ref{tab:t2} shows the impact of the various components of the 3D Prior Aided Audio Disentanglement module.
The entry labelled 'use ent. aud.' presents the experimental results obtained by using the entire speech audio. The entries labelled 'use spe. aud.' and 'use dis. aud.' present the experimental results obtained by using only the features related to 3D awarded speech movements \(f_{exp-aud}\) and using the two disentanglement features, respectively.

\begin{table}[t]
\resizebox{\linewidth}{!}{
\begin{tabular}{cccccc}
\toprule
\multicolumn{1}{c|}{Method}     & SSIM↑          & LPIPS↓         & AU Acc↑        & LMD-79↓           & SyncNet↑       \\ \midrule \midrule
\multicolumn{1}{c|}{Ground-truth}      & -              & -              & -              & -              & 2.279          \\ 
\multicolumn{1}{c|}{w/o Global}       & 0.819          & 0.041          & 72.4\%          & 2.637          & 1.721          \\
\multicolumn{1}{c|}{w/o AU}                & 0.832          & 0.039 & 73.1\%          & 2.624          & 1.719          \\
\multicolumn{1}{c|}{Ours}         & \textbf{0.835} & \textbf{0.037} & \textbf{75.3\%} & \textbf{2.529} & \textbf{1.956} \\  \bottomrule
\end{tabular}}
\caption{The ablation study of Standardized Space module.}
\vspace{-20pt}
\label{tab:t3}
\end{table}

The impact of various components of the Standardized Space is presented in Table \ref{tab:t3}.
The entries labelled "w/o Global" and "w/o AU" indicate the experimental results after removing the global space and the local AU semantic space, respectively. 
The table demonstrates that removing these components step by step results in a decrease in the quality of the generated images and the lip synchronization. 


\section{Conclusion}

This paper presents NeRF-3DTalker to synthesize 3D talking head with free views.
To tackle the problem of blurry novel view facial images, we parameterize the face using 3D priors extracted by 3DMM.
Furthermore, we introduce a 3D Prior Aided Audio Disentanglement module to simplify the learning process of the NeRF and 
align the acoustic space with the 3D visual space, thereby enhancing lip synchronization.  Subsequently, those 3D priors and disentangled semantic features are utilized to guide the presented conditional NeRF in synthesizing 3D talking heads.
Finally, a Standardized Space is designed to bring the generated frames distant from the speaker’s motion space back to the real space from both global and local semantic perspectives. 
Extensive qualitative and quantitative experiments demonstrate the superiority of our NeRF-3DTalker.

\textbf{Acknowledgement.} This work was supported by National Natural Science Foundation of China under Grant No. 62172294.


\bibliographystyle{IEEEtran}
\bibliography{IEEEfull}

\begin{thebibliography}{10}
\providecommand{\url}[1]{#1}
\csname url@samestyle\endcsname
\providecommand{\newblock}{\relax}
\providecommand{\bibinfo}[2]{#2}
\providecommand{\BIBentrySTDinterwordspacing}{\spaceskip=0pt\relax}
\providecommand{\BIBentryALTinterwordstretchfactor}{4}
\providecommand{\BIBentryALTinterwordspacing}{\spaceskip=\fontdimen2\font plus
\BIBentryALTinterwordstretchfactor\fontdimen3\font minus \fontdimen4\font\relax}
\providecommand{\BIBforeignlanguage}[2]{{%
\expandafter\ifx\csname l@#1\endcsname\relax
\typeout{** WARNING: IEEEtran.bst: No hyphenation pattern has been}%
\typeout{** loaded for the language `#1'. Using the pattern for}%
\typeout{** the default language instead.}%
\else
\language=\csname l@#1\endcsname
\fi
#2}}
\providecommand{\BIBdecl}{\relax}
\BIBdecl

\bibitem{ding2017exprgan}
H.~Ding, K.~Sricharan, and R.~Chellappa, ``Exprgan: Facial expression editing with controllable expression intensity,'' 2017.

\bibitem{pumarola2018ganimation}
A.~Pumarola, A.~Agudo, A.~M. Martinez, A.~Sanfeliu, and F.~Moreno-Noguer, ``Ganimation: Anatomically-aware facial animation from a single image,'' in \emph{Proceedings of the European conference on computer vision (ECCV)}, 2018, pp. 818--833.

\bibitem{ji2021audio}
X.~Ji, H.~Zhou, K.~Wang, W.~Wu, C.~C. Loy, X.~Cao, and F.~Xu, ``Audio-driven emotional video portraits,'' in \emph{Proceedings of the IEEE/CVF conference on computer vision and pattern recognition}, 2021, pp. 14\,080--14\,089.

\bibitem{lu2021live}
Y.~Lu, J.~Chai, and X.~Cao, ``Live speech portraits: real-time photorealistic talking-head animation,'' \emph{ACM Transactions on Graphics (TOG)}, vol.~40, no.~6, pp. 1--17, 2021.

\bibitem{thies2020neural}
J.~Thies, M.~Elgharib, A.~Tewari, C.~Theobalt, and M.~Nie{\ss}ner, ``Neural voice puppetry: Audio-driven facial reenactment,'' in \emph{Computer Vision--ECCV 2020: 16th European Conference, Glasgow, UK, August 23--28, 2020, Proceedings, Part XVI 16}.\hskip 1em plus 0.5em minus 0.4em\relax Springer, 2020, pp. 716--731.

\bibitem{chen2021talking}
S.~Chen, Z.~Liu, J.~Liu, Z.~Yan, and L.~Wang, ``Talking head generation with audio and speech related facial action units,'' in \emph{32nd British Machine Vision Conference (BMVC)}, 2021, p. 353.

\bibitem{zhua2023audio}
Y.~Zhua, C.~Zhanga, Q.~Liub, and X.~Zhoub, ``Audio-driven talking head video generation with diffusion model,'' in \emph{ICASSP 2023-2023 IEEE International Conference on Acoustics, Speech and Signal Processing (ICASSP)}.\hskip 1em plus 0.5em minus 0.4em\relax IEEE, 2023, pp. 1--5.

\bibitem{shen2022learning}
S.~Shen, W.~Li, Z.~Zhu, Y.~Duan, J.~Zhou, and J.~Lu, ``Learning dynamic facial radiance fields for few-shot talking head synthesis,'' in \emph{Computer Vision--ECCV 2022: 17th European Conference, Tel Aviv, Israel, October 23--27, 2022, Proceedings, Part XII}.\hskip 1em plus 0.5em minus 0.4em\relax Springer, 2022, pp. 666--682.

\bibitem{stypulkowski2023diffused}
M.~Stypu{\l}kowski, K.~Vougioukas, S.~He, M.~Zi{\k{e}}ba, S.~Petridis, and M.~Pantic, ``Diffused heads: Diffusion models beat gans on talking-face generation,'' \emph{arXiv preprint arXiv:2301.03396}, 2023.

\bibitem{multomm}
Z.~Liu, X.~Liu, S.~Chen, J.~Liu, L.~Wang, and C.~Bi, ``Multimodal fusion for talking face generation utilizing speech-related facial action units,'' \emph{ACM Trans. Multimedia Comput. Commun. Appl.}, vol.~20, no.~9, Sep. 2024.

\bibitem{2020NeRF}
B.~Mildenhall, P.~P. Srinivasan, M.~Tancik, J.~T. Barron, R.~Ramamoorthi, and N.~Ren, ``Nerf: Representing scenes as neural radiance fields for view synthesis,'' 2020.

\bibitem{bi2024nerf}
C.~Bi, X.~Liu, and Z.~Liu, ``Nerf-ad: Neural radiance field with attention-based disentanglement for talking face synthesis,'' in \emph{ICASSP 2024-2024 IEEE International Conference on Acoustics, Speech and Signal Processing (ICASSP)}.\hskip 1em plus 0.5em minus 0.4em\relax IEEE, 2024, pp. 3490--3494.

\bibitem{tran2018nonlinear}
L.~Tran and X.~Liu, ``Nonlinear 3d face morphable model,'' in \emph{Proceedings of the IEEE conference on computer vision and pattern recognition}, 2018, pp. 7346--7355.

\bibitem{yao2022dfa}
S.~Yao, R.~Zhong, Y.~Yan, G.~Zhai, and X.~Yang, ``Dfa-nerf: personalized talking head generation via disentangled face attributes neural rendering,'' \emph{arXiv preprint arXiv:2201.00791}, 2022.

\bibitem{van2017neural}
A.~Van Den~Oord, O.~Vinyals \emph{et~al.}, ``Neural discrete representation learning,'' \emph{Advances in neural information processing systems}, vol.~30, 2017.

\bibitem{egger20203d}
B.~Egger, W.~A. Smith, A.~Tewari, S.~Wuhrer, M.~Zollhoefer, T.~Beeler, F.~Bernard, T.~Bolkart, A.~Kortylewski, S.~Romdhani \emph{et~al.}, ``3d morphable face models—past, present, and future,'' \emph{ACM Transactions on Graphics (ToG)}, vol.~39, no.~5, pp. 1--38, 2020.

\bibitem{hong2022headnerf}
Y.~Hong, B.~Peng, H.~Xiao, L.~Liu, and J.~Zhang, ``Headnerf: A real-time nerf-based parametric head model,'' in \emph{Proceedings of the IEEE/CVF Conference on Computer Vision and Pattern Recognition}, 2022, pp. 20\,374--20\,384.

\bibitem{zhang2022sadtalker}
W.~Zhang, X.~Cun, X.~Wang, Y.~Zhang, X.~Shen, Y.~Guo, Y.~Shan, and F.~Wang, ``Sadtalker: Learning realistic 3d motion coefficients for stylized audio-driven single image talking face animation,'' \emph{arXiv preprint arXiv:2211.12194}, 2022.

\bibitem{prajwal2020lip}
K.~Prajwal, R.~Mukhopadhyay, V.~P. Namboodiri, and C.~Jawahar, ``A lip sync expert is all you need for speech to lip generation in the wild,'' in \emph{Proceedings of the 28th ACM International Conference on Multimedia}, 2020, pp. 484--492.

\bibitem{niemeyer2021giraffe}
M.~Niemeyer and A.~Geiger, ``Giraffe: Representing scenes as compositional generative neural feature fields,'' in \emph{Proceedings of the IEEE/CVF Conference on Computer Vision and Pattern Recognition}, 2021, pp. 11\,453--11\,464.

\bibitem{karras2020analyzing}
T.~Karras, S.~Laine, M.~Aittala, J.~Hellsten, J.~Lehtinen, and T.~Aila, ``Analyzing and improving the image quality of stylegan,'' in \emph{Proceedings of the IEEE/CVF conference on computer vision and pattern recognition}, 2020, pp. 8110--8119.

\bibitem{baltruvsaitis2015cross}
T.~Baltru{\v{s}}aitis, M.~Mahmoud, and P.~Robinson, ``Cross-dataset learning and person-specific normalisation for automatic action unit detection,'' in \emph{2015 11th IEEE International Conference and Workshops on Automatic Face and Gesture Recognition (FG)}, vol.~6.\hskip 1em plus 0.5em minus 0.4em\relax IEEE, 2015, pp. 1--6.

\bibitem{baltrusaitis2018openface}
T.~Baltrusaitis, A.~Zadeh, Y.~C. Lim, and L.-P. Morency, ``Openface 2.0: Facial behavior analysis toolkit,'' in \emph{2018 13th IEEE international conference on automatic face \& gesture recognition (FG 2018)}.\hskip 1em plus 0.5em minus 0.4em\relax IEEE, 2018, pp. 59--66.

\bibitem{milletari2016v}
F.~Milletari, N.~Navab, and S.-A. Ahmadi, ``V-net: Fully convolutional neural networks for volumetric medical image segmentation,'' in \emph{2016 fourth international conference on 3D vision (3DV)}.\hskip 1em plus 0.5em minus 0.4em\relax Ieee, 2016, pp. 565--571.

\bibitem{xing2023codetalker}
J.~Xing, M.~Xia, Y.~Zhang, X.~Cun, J.~Wang, and T.-T. Wong, ``Codetalker: Speech-driven 3d facial animation with discrete motion prior,'' in \emph{Proceedings of the IEEE/CVF Conference on Computer Vision and Pattern Recognition}, 2023, pp. 12\,780--12\,790.

\bibitem{ye2023geneface}
Z.~Ye, Z.~Jiang, Y.~Ren, J.~Liu, J.~He, and Z.~Zhao, ``Geneface: Generalized and high-fidelity audio-driven 3d talking face synthesis,'' in \emph{The Eleventh International Conference on Learning Representations, {ICLR} 2023, Kigali, Rwanda, May 1-5, 2023}.

\bibitem{synctalk}
Z.~Peng, W.~Hu, Y.~Shi, X.~Zhu, X.~Zhang, H.~Zhao, J.~He, H.~Liu, and Z.~Fan, ``Synctalk: The devil is in the synchronization for talking head synthesis,'' in \emph{Proceedings of the IEEE/CVF Conference on Computer Vision and Pattern Recognition}, 2024, pp. 666--676.

\bibitem{wang2004image}
Z.~Wang, A.~C. Bovik, H.~R. Sheikh, and E.~P. Simoncelli, ``Image quality assessment: from error visibility to structural similarity,'' \emph{IEEE transactions on image processing}, vol.~13, no.~4, pp. 600--612, 2004.

\bibitem{zhang2018unreasonable}
R.~Zhang, P.~Isola, A.~A. Efros, E.~Shechtman, and O.~Wang, ``The unreasonable effectiveness of deep features as a perceptual metric,'' in \emph{Proceedings of the IEEE conference on computer vision and pattern recognition}, 2018, pp. 586--595.

\bibitem{chen2018lip}
L.~Chen, Z.~Li, R.~K. Maddox, Z.~Duan, and C.~Xu, ``Lip movements generation at a glance,'' in \emph{Proceedings of the European conference on computer vision (ECCV)}, 2018, pp. 520--535.

\end{thebibliography}

\end{document}